\documentclass[manuscript]{aastex}
\shorttitle{magnetar model of GRB 130831A}
\shortauthors{Zhang et al.}

\begin{document}
\title{Modeling the Multi-Band Afterglow of GRB 130831A: \\ Evidence for a Spinning-Down Magnetar Dominated by Gravitational Wave Losses?}
\author{Q. Zhang\altaffilmark{1,2}, Y. F. Huang\altaffilmark{3,4}, H. S. Zong\altaffilmark{1,2,5}}
\altaffiltext{1}{School of Physics, Nanjing University, Nanjing 210093, China; zonghs@nju.edu.cn}
\altaffiltext{2}{Joint Center for Particle, Nuclear Physics and Cosmology, Nanjing 210093, China}
\altaffiltext{3}{School of Astronomy and Space Science, Nanjing University, Nanjing 210093, China; hyf@nju.edu.cn }
\altaffiltext{4}{Key Laboratory of Modern Astronomy and Astrophysics (Nanjing University), Ministry of
Education, Nanjing 210093, China}
\altaffiltext{5}{State Key Laboratory of Theoretical Physics, Institute of Theoretical Physics, CAS, Beijing, 100190, China}

\begin{abstract}
The X-ray afterglow of GRB 130831A shows an ``internal plateau'' with a decay slope of $\sim$ 0.8, followed by a steep drop at around $10^5$ s with a slope of $\sim$ 6. After the drop, the X-ray afterglow continues with a much shallower decay.
The optical afterglow exhibits two segments of plateaus separated by a luminous optical flare, followed by a normal decay with a slope basically consistent with that of the late-time  X-ray afterglow.  The decay of the internal X-ray plateau is much steeper than what we expect in the simplest magnetar model. We propose a scenario in which the magnetar undergoes gravitational-wave-driven r-mode instability, and the spin-down is  dominated by gravitational wave losses up to the end of the steep plateau, so that such a relatively steep plateau can be interpreted as  the internal emission of the magnetar wind and the sharp drop can be produced when the magnetar collapses into a black hole. This scenario also predicts an initial X-ray  plateau lasting for hundreds of seconds with an approximately constant flux which is compatible with observation.
Assuming that the magnetar wind has a negligible contribution in the optical band, we interpret the optical afterglow as the forward shock emission by invoking the energy injection from a continuously refreshed shock following the prompt emission phase. It is shown that our model can basically describe
the temporal evolution of the multi-band afterglow of GRB 130831A.

\end{abstract}


\section{INTRODUCTION}\label{intro}
Gamma-ray bursts (GRBs) are the most energetic stellar explosions in the universe. These events produce a short prompt $\gamma$-ray emission followed by a multi-band afterglow that can be observed up to several years. The afterglows of GRBs are thought to originate from the synchrotron emission of shock-accelerated electrons produced by the interaction between the outflow and the external medium \citep{Rees92,Mes93,Mes97,Sari98,Huang99,Cheva00}. Before the launch of the {\it Swift} satellite \citep{Geh04}, the afterglows were often observed to decay as a power law with time ($F_{\nu}\propto t^{-1}-t^{-2}$) and have a power-law spectrum ($F_{\nu}\propto \nu^{-0.9\pm0.5}$) which can be well explained by the synchrotron radiation from the external forward shock \citep{Kumar15}. However,
{\it Swift} observations have shown evidence of more complex phenomena  during the afterglow phase, such as X-ray  flares \citep{Fal07,Chin10,Mar11} and the shallow decay phase \citep[or the so-called ``plateau'';][]{Nou06,OBrien06,Zhang06,Evans09,Marg13}, which challenge the standard afterglow models \citep{Sari98,Cheva00}.

A good fraction of {\it Swift} GRB afterglows exhibit an X-ray plateau that lasts for  $10^3-10^4$ s with a slope of $\sim$ 0.3, followed by a normal or steeper decay with a slope of $\sim 1-2$ \citep{Nou06,Zhang06,Evans09,Rac09,Grup13,Marg13}. Since this type of plateau can be  typically interpreted as afterglow emission from the external shock in the context of the energy injection model \citep{Zhang06}, it is often referred to as an ``external plateau'', though some internal dissipation processes \citep[e.g.,][]{Ghi07} may also account for this phenomenon. In a small subset of {\it Swift} GRB afterglows, observations have shown an X-ray plateau followed by a very sharp drop with a decay slope steeper than 3, sometimes approaching $\sim 9-10$ \citep{Liang07,Troja07,Lyons10,Row10,Row13,Lu14,Lu15}.
Such a steep decay cannot be accommodated in any external forward shock model\footnote{From this point onwards, we use ``external shock''
as a synonym of ``forward shock'', and we do not consider reverse shock emission.}, and the entire plateau emission is usually attributed to the internal dissipation of a central engine wind. Clues for such ``internal plateaus'' have been found in several short GRBs \citep{Row13}, but are indicated in only a small portion of long GRBs \citep{Troja07,Lyons10}.
 Two types of GRB central engines  have been widely studied in the literature: a hyper-accreting stellar-mass black hole \citep[e.g.,][]{Pop99,Nara01,Lei13}, and a rapidly spinning, strongly magnetized neutron star or ``millisecond magnetar'' \citep{Usov92,Thom94,Dai98,Whee00,Zhang01,Metz08,Bucc12}.
Within the scenario of a millisecond magnetar, the internal plateau can be interpreted as the internal emission of a spinning-down magnetar which collapses into a black hole at the end of the plateau \citep{Troja07,Row10,Zhang14}. Since the GRB outflow still produces X-ray afterglow by the external shock during the internal plateau phase, it is expected to emerge once the X-ray emission from the magnetar wind drops below the external component. So far, this has been seen clearly in the X-ray afterglow of the long GRB 070110 \citep{Troja07}.

GRB 130831A \citep{De16} is another event that exhibits a superposition of external
and internal emission in the X-ray afterglow. The X-ray light curve (LC) shows a
shallow decay with a slope of $\sim$ 0.8, followed by a sharp drop
at about $10^5$ s with a slope of $\sim$ 6. After the drop, the X-ray
afterglow continues with a much shallower decay. Such a steep drop indicates that the
shallow decay phase must be of ``internal origin'' \citep{De16}. However, the slope of
this internal plateau is much steeper than the usually observed, and cannot be explained
by the simplest magnetar model which predicts a plateau with approximately constant
flux. \citet{De16} proposed  a more elaborate model of the magnetar spin-down, in which the magnetic field is assumed to decay as the spin period
increases \citep{Metz11}. This is possible but detailed numerical calculations are
needed to test this possibility.

In this paper, we propose a new scenario to explain the internal X-ray  plateau of GRB 130831A. In our model, the nascent magnetar  could undergo some
nonaxisymmetric stellar instabilities \citep[e.g., r-mode instability;][]{and98,Fri98}, leading to strong gravitational wave (GW) losses which  would affect the magnetar's spin-down \citep{Corsi09}. In the context of r-mode instability \citep{Sa05,Sa06,Yu10}, if the gravitational braking timescale ($\tau_{\rm g}$) is much smaller than the magnetic braking timescale ($\tau_{\rm m}$), the spin-down would be first dominated by GW losses and the luminosity of the magnetar wind can be regarded as a constant within a timescale comparable to $\tau_{\rm g}$ \citep[$\sim$ a few $\times10^2$ s for a neutron rotating with the Keplerian frequency;][]{Sa05}. After that, the spin of the magnetar would be remarkably decelerated and the luminosity of the wind evolves as $L_{\rm w}\propto t^{-4/5}$. At later times, the spin-down would be eventually dominated by the magnetic dipole radiation and $L_{\rm w}$ decays as $t^{-2}$. During the spin-down process, the magnetar could collapse into a black hole and a steep drop of the wind luminosity would be observed.
We test this scenario using the multi-band afterglow of GRB 130831A in this work.

In addition, GRB 130831A has a well-sampled and multi-band optical afterglow. The optical LCs exhibit two segments of plateaus separated by a luminous optical flare, followed by a normal decay with a slope basically consistent with that of the late-time  X-ray afterglow \citep{De16}. \citet{De16} interpreted the afterglow after the flare  as the forward shock (FS) emission in the context of the standard afterglow models by requiring the ejecta decelerating at $\sim$ 4ks. However, their model cannot explain the initial optical plateau before $\sim$ 500 s. To solve this problem, we assume that these two plateaus have the same FS origin and are produced by an energy injection process \citep{Zhang06}. We will show in Section \ref{model} that this energy injection cannot be supplied by the magnetic wind \citep{Dai98,Zhang01}, since the wind luminosity has a negligible contribution to the FS emission. We also assume that the magnetar wind emission contributes negligibly in the optical band and we invoke the model of \citet{Sari00} to explain the entire optical afterglow. In this case, the energy injection is produced by a continuously refreshed shock following the prompt emission phase \citep{Rees98}.

Our paper is organized as follows.  We summarize the observational facts of GRB 130831A and refit the afterglow LCs in Section \ref{obs}. In Section \ref{model}, we model the external  afterglow by invoking the energy injection model, and interpret the internal X-ray afterglow as the magnetar wind emission produced by the spin-down process dominated by GW  losses. In Section \ref{fitting}, we compare our theoretical afterglow LCs with observations. Finally, we present our conclusions and give a brief discussion in Section \ref{conclusion}. Throughout the paper, the convention $F_{\nu}\propto\nu^{-\beta}t^{-\alpha}$ is followed, and we use the standard notation $Q_x=Q/10^{x}$ with $Q$ being a generic quantity in cgs units. We assume a concordance cosmology with $H_0=70 ~\rm{km}~ \rm{s}^{-1} \rm{Mpc}^{-1}$, $\Omega_{\rm{M}}=0.27$ and $\Omega_\Lambda=0.73$ \citep{Jar11}.
All the errors are given at the $1\sigma$ confidence level (CL).

\section{OBSERVATIONAL FACTS}\label{obs}
GRB 130831A triggered the {\it Swift} Burst Alert Telescope \citep[BAT;][]{Bart05} at $T_0=$13:04:16.54 UT on 2013 August 31 and was also observed by {\it Konus-Wind} onboard the {\it WIND} spacecraft. The light curve in the 15 -- 350 keV energy range shows a main pulse with a fast rise and exponential decay (FRED) shape, followed by some extended emission that lasts until $T_0+41$ s. The measured duration of $T_{90}$ in the 15 -- 350~keV band is $30.2\pm1.4$ s \citep{De16}. The time-averaged spectrum between 20 keV and 15 MeV can be fitted by a Band function \citep{Band93} with the peak energy $E_{\rm{p}}=55\pm4$ keV, the low-energy photon index $\alpha=-0.61\pm0.06$ and the high-energy photon index $\beta=-2.3\pm0.3$. The fluence between 20 keV and 10 MeV is
$(7.6\pm0.2)\times10^{-6}$ ${\rm erg\ cm^{-2}}$ \citep{Gol13}. With a redshift of $z=0.479$ \citep{Cucc13}, this corresponds to an isotropically equivalent energy $E_{\gamma}=1.06\times10^{52}$ erg in the 1 -- 10000 keV rest-frame energy band \citep{De16}.

\subsection{X-Ray Afterglow} \label{Xobs}
\begin{figure}
\includegraphics[scale=.50]{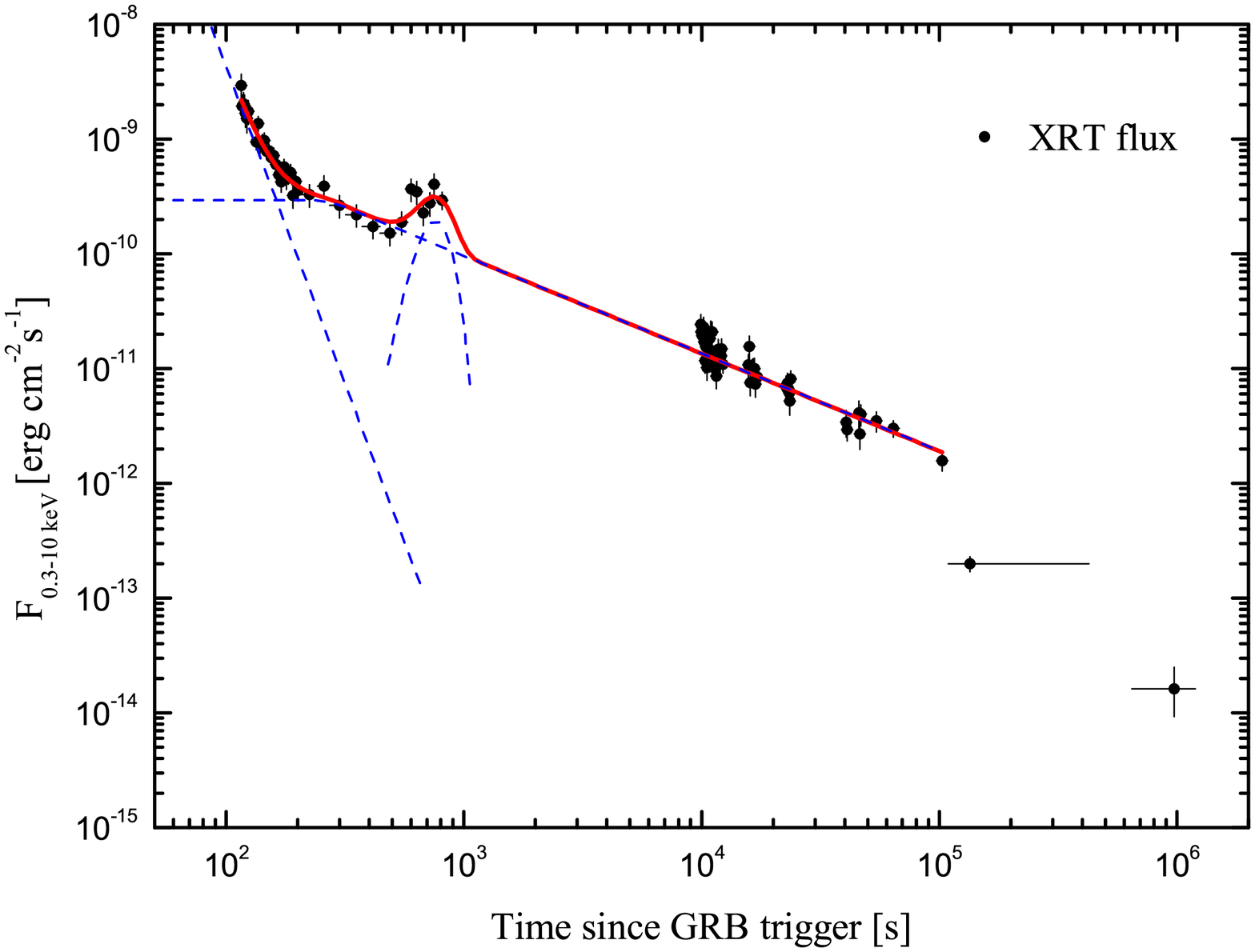}
\caption{Fit of the X-ray flux of GRB 130831A with the PL+BPL+Gauss model (solid line). The data are taken from http://www.swift.ac.uk/burst\_analyser/00568849/ \citep{Evans09}. The three fitting components (dashed lines) are displayed for clarity. \label{Xfit1}}
\end{figure}

The X-ray Telescope \citep[XRT;][]{Burr05} began observations of GRB 130831A 125.8 s after the BAT trigger, and monitored the source until 2013 September 14.
The X-ray afterglow of GRB 130831A was also observed by {\it Chandra} at $T_0+16.6$ d and $T_0+33.1$ d \citep{De16}. As shown in Figure \ref{Xfit1}, the X-ray LC shows an initial fast decay ending at about 200 s, then it gives way to a shallower decay up to about 100 ks. After that, the flux drops quickly, then evolves slowly again at around 200 ks. In addition, there is a X-ray flare between about 500 and 900 s.

\citet{De16} fit the X-ray afterglow with the sum of an initial power law (PL),
a broken power law (BPL) and a final PL together with a Gaussian function. They found that
the plateau between 0.3 and 100 ks can be well described by a single PL with
slope of $\sim$ 0.8.
It is not clear how this plateau evolves before 0.3 ks. According to our GW radiation
dominated magnetar model, there should be an initial plateau within hundreds of seconds.
However, the X-ray data show no evidence of such an initial plateau. One
possibility is that this short plateau indeed exists but ends before $\sim$ 300 s
and is buried by the initial steep decay of the X-ray afterglow.
Therefore, we give a physically motivated fit to the X-ray afterglow based on our
new considerations. We use the data of 0.3 -- 10 keV unabsorbed X-ray flux produced
by an automatic analysis procedure \citep{Evans07,Evans09}.
We focus on the data before $\sim10^5$ s and fit the LC with an initial PL, a BPL and
a Gaussian function. We fix the initial decay index of the BPL as zero which is
predicted by our magnetar model. The main purpose of our fit here is to draw some key information from the observational
data, such as the time of light curve breaks, the flux level, the power-law indices, etc.
We use the Origin software (Version 8.5\footnote{www.originlab.com}) to perform a least-square fit to obtain the model parameters.
Figure \ref{Xfit1} shows our fit of
the X-ray afterglow before $10^5$ s. The main fitting parameters are listed in Table~\ref{table1}.
We note that our parameter values are basically consistent with those given by \citet{De16} except that we have two
extra parameters, i.e., the flux and the break time ($\sim$ 270 s) of the initial short plateau. The main difference between
our results and those of \citet{De16} is
that we have introduced an initial plateau in our fit, though such a short plateau cannot be identified just from the data.

For late X-ray afterglow after $\sim$ 100 ks, we adopt the results of \citet{De16}: the best-fitting slope of
the steep drop is $6.8^{+2.0}_{-1.5}$ and the late shallow decay index is $1.11^{+0.22}_{-0.29}$. In addition,
the X-ray spectrum between 9 and 132 ks can be modeled with an absorbed PL, which yields the
spectral index $\beta_{\rm X}=0.77\pm0.07$ and the host absorbing column density $N_{\rm H}=6.8^{+3.3}_{-3.1}\times10^{20} {\rm cm}^{-2}$ \citep{De16}.

\subsection{Optical Afterglow}
The optical afterglow of GRB 130831A was  observed by {\it Swift}/UVOT \citep{Rom05} from 114 s after the trigger \citep{Hagen13}, and it was also monitored by several ground telescopes and IKI Network for Transients, such as the Reionization and Transients Infra-Red camera \citep[RATIR;][]{But12}, Skynet \citep{Rei05} and the International Scientific Optical-Observation Network \citep[ISON;][]{Mol08,Poz13}. \citet{De16} gave a detailed study of the multi-color afterglow  ranging from the infrared to ultraviolet bands, while the data analysis of SN 2013fu that is associated with this burst can be found in \citet{Cano14}.

The optical LCs show an initial short plateau (in the $U$ band) that lasts until $\sim$ 500 s, followed by a bump peaking at around 730 s. This optical bump is basically concurrent with the X-ray flare and evolves rapidly. After the bump, there is another plateau that in turn gives way to  a steeper decay at $\sim$ 5 ks \citep{De16}. \citet{De16} fit the LCs between 3.5 and 15 ks with a BPL.
They assumed that the observed bumps at $\sim$ 730 s are optical flares since they peak at the same time as the X-ray flare and have a rapid temporal behavior,
while they have not taken into account the early data (before $\sim$ 3.5 ks) in their fit.

\begin{figure}
\includegraphics[scale=.50]{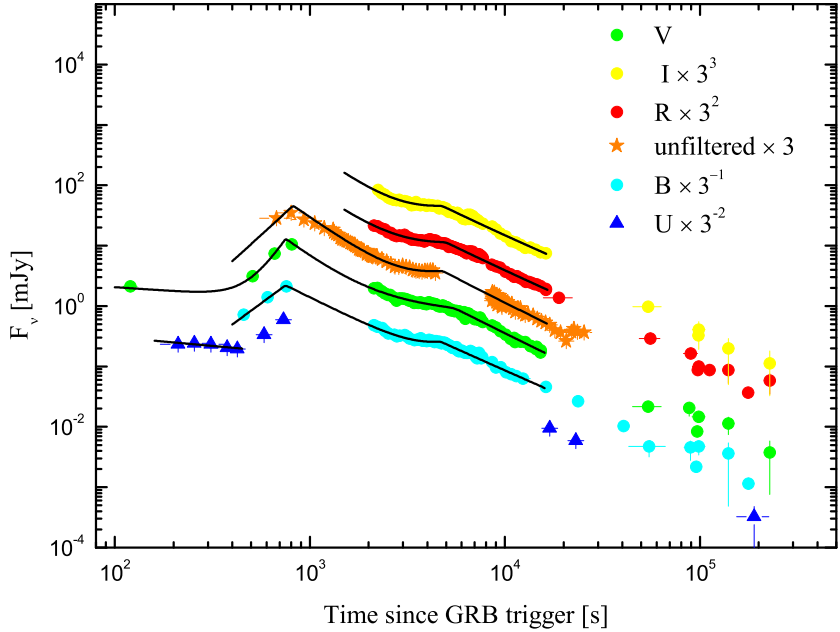}
\caption{Fit of the optical light curves of GRB 130831A before 16 ks (solid lines). The $V, B$ and unfiltered bands are fitted with two BPLs, while the $R$ and $I$ bands are fitted with a PL+BPL. For the $U$ band, we fit only the initial plateau with a PL. The data are taken from \citet{De16}.
Adjacent light curves have been offset by a factor of three for clarity, with $V$ band unscaled.\label{optfit1}}
\end{figure}

Our main interest here is the origin of the initial optical plateau.
As stated in Section \ref{intro}, it is natural to assume the initial plateau is
produced by the external shock, and has the same origin as the later
plateau (between 3.5 and 5 ks). In this case, a fit including the early data (especially for
the $V$ band) is necessary. Therefore, we refit the optical afterglow before $\sim$ 16 ks (Figure \ref{optfit1}).
We fit the $V, B$ and unfiltered bands with two BPLs. For
the $R$ and $I$ bands, we fit the decaying segment of the flare with a PL and fit the post-flare data with
a BPL. For the $U$ band,
we fit only the initial plateau with a PL. We perform the same fitting procedure as that in Section \ref{Xobs}.
As seen from Table \ref{table2}, our fit gives relatively high $\chi^2/dof$.
Such high $\chi^2/dof$ are also given by \citet{De16}, they are mainly due to
some ``wiggles'' of the densely sampled LCs in this phase. Comparing the parameters with
those of \citet{De16}, we have similar break times at $\sim$ 5 ks, while the decay
indices ($\sim$ 1.32 -- 1.55) are slightly shallower than theirs ($\sim$ 1.45 -- 1.82).  However, the biggest
difference lies in the decay slopes of the plateaus. For the $V$ and $U$ bands, our
obtained slopes ($\sim$ 0.2) are significantly smaller than theirs ($\sim$ 0.8), since we
have included the early data before $\sim$ 3 ks. While for other bands, we even give
negative slopes. It is not unexpected since the superposition of
a decaying optical flare and a rising peak can produce a plateau-like feature. This is consistent with the model of \citet{De16}, in which the late plateau results from the combination of the decaying optical flare and the rising of the FS peak. However, their model cannot explain the initial short plateau.
Since we assume these two plateaus are of the same origin, a plateau slope of $\sim$ 0.2 is preferred. For the post-plateau decay
slope, we use the value of $\sim$ 1.59 given by \citet{De16} since they made a more in-depth analysis in this phase.
In addition, The spectral energy distribution (SED) with the optical and X-ray data at 2 days gives a spectral index $\beta_{\rm OX}=1.03^{+0.05}_{-0.04}$ and a small or absent amount of host-galaxy reddening $E(B-V)=0.02\pm0.01$ mag \citep{De16,Cano14}.

\section{MODEL}\label{model}

\subsection{Afterglow of External Origin}\label{external}

The optical afterglow of GRB 130831A shows a long plateau before $\sim$ 5 ks, a similar plateau must have appeared in the external component of the X-ray afterglow in this period if we assume the X-ray and the optical emissions lie on the same spectral segment throughout the afterglow observations.
Theoretically, several models have been proposed to interpret the afterglow plateau \citep[e.g.,][]{Zhang01,Sari00,Eich06,Ioka06,Ghi07,Shao07,Uhm07}, some of them can even explain the chromatic breaks in GRB afterglows (Ghisellini et al. 2007; Uhm \& Beloborodov 2007).
For the simplest case that the X-ray and the optical LCs have achromatic behaviors, the energy injection models are usually adopted and the plateau phase can be produced by a significant continuous energy injection into the decelerating forward shock. There are at least two possible physical origins for the energy injection, depending on whether the central engine is long or short lived \citep{Zhang06}: (i) The central engine is long lived and the luminosity is assumed to evolve as $L(t)\propto t^{-q}$ with $q<1$. A specific case, corresponding to $q=0$, is that the energy injection is from the spin-down of a millisecond pulsar or magnetar \citep{Dai98,Zhang01}. (ii) The engine is short lived (i.e., with duration comparable to that of the prompt phase) and is assumed to produce shells with a steep power-law distribution of Lorentz factors and most of the system's energy is carried by the slower material \citep{Rees98,Sari00}. Both scenarios interpret the plateau as afterglow emission from a continuous refreshed shock, which requires the injected energy largely exceeding the initial kinetic energy of the ejecta.

For GRB 130831A, the possibility that the energy injection is from the spin-down of a magnetar can be excluded. To show that, we give a rough estimate. The initial plateau luminosity of the X-ray afterglow is $\sim10^{47}$ erg s$^{-1}$ (see table \ref{table1}). We assume that the efficiency of converting the magnetic dipole emission into X-ray radiation is $\sim$ 0.1, then the isotropically equivalent luminosity of the magnetar wind is $\sim10^{48}$ erg s$^{-1}$.  The injected energy increases as $E_{\rm{inj}}\propto t^{1-q}$ \citep{Zhang06}, then the energy injection before $\sim$ 300 s is $\sim10^{50}$ erg, and we would see an increase in the injected energy by a factor of $\sim$ 3 between 300 and $10^5$ s. That is, the energy injection from the magnetar wind  before $10^5$ s is at most $10^{51}$ erg, which is much less than the isotropically equivalent energy of the prompt emission and cannot exceed the initial kinetic energy of the fireball. Therefore, the injected kinetic energy from the magnetic dipole emission is basically negligible.
In the following calculations, we use the energy injection model proposed by \citet{Sari00} to explain the external plateau of GRB 130831A.

According to \citet{Sari00}, the ejecta are assumed to possess a continuous distribution of Lorentz factors, e.g., the amount of ejected
mass moving with Lorentz factors greater than $\gamma$ is
\begin{equation}
M\left(>\gamma\right)\propto\gamma^{-s},
\end{equation}
down to some minimum Lorentz factor $\gamma_{\rm min}\le\gamma_0$, where the initial Lorentz factor $\gamma_0$ is assumed to be at the deceleration time. $s>1$ is required to have a change in the fireball dynamics.
The energy of the blastwave increases as $E\propto\gamma^{\left(1-s\right)}$, and its Lorentz factor $\gamma\propto t^{\left[-3/\left(7+s\right)\right]}$ for a homogeneous interstellar medium (ISM) environment which is the case for GRB 130831A (see below).

The evolution of $E$ can be written as \citep{Laskar15}
\begin{equation}\label{Et}
E\left(t\right)=\left\{
\begin{array}{ll}
E_{\rm i}=E_{\rm f}\left(\frac{t_{\rm i}}{t_{\rm f}}\right)^m, & t\le t_{\rm i},\\
E_{\rm f}\left(\frac{t}{t_{\rm f}}\right)^m, & t_{\rm i}<t<t_{\rm f},\\
E_{\rm f}, & t\ge t_{\rm f},
\end{array}
\right.
\end{equation}
where $m=3\left(s-1\right)/\left(7+s\right)$ for the ISM case,  $t_{\rm i}$ and $t_{\rm f}$ are the start and end time of the energy injection, respectively;  while $E_{\rm i}$ and $E_{\rm f}$ are the initial and final energy of the blastwave, respectively.
For GRB 130831A, the optical plateau starts at around 120 s, so we take $t_{\rm i}=100$ s, which is also assumed to be the deceleration time, and take
$t_{\rm f}=5000$ s.

In the following calculations, we adopt $\alpha_1=0.2$ for the plateau decay slope, $\alpha_2=1.59$ for the post-plateau slope and $\beta_{\rm OX}=1.03$ for the spectral index. The closure relations of the post-plateau phase require an ISM environment. The observing frequency should be between the synchrotron peak frequency $\nu_{\rm m}$ and the cooling frequency $\nu_{\rm c}$, i.e., $\nu_{\rm m}<\nu_{\rm opt}<\nu_{\rm X}<\nu_{\rm c}$, and the electron spectral index is derived as $p=3.06$ \citep{De16}. Using the closure relations of the plateau phase \citep{Sari00}, we obtain $s=4.4$, and $m=0.9$.

We follow the formalism of \citet{Gao13} to compute the break frequencies and the peak flux ($\nu_{\rm m}$, $\nu_{\rm c}$ and $F_{\nu,\rm max}$) but replace the blastwave energy with Equation (\ref{Et}) and consider the synchrotron self-Compton (SSC) effect \citep{Sari01}.
For  $\nu_{\rm m}<\nu<\nu_{\rm c}$, the flux density is
\begin{equation}
F_{\nu}=3.5~\mu {\rm{Jy}} E^{1.52}_{52}\epsilon_{\rm e,-1}^{2.06}\epsilon_{\rm B,-2}^{1.02}n_{0}^{1/2} t_{5}^{-1.55}\left(\frac{\nu}{\nu_{\rm R}}\right)^{-1.03},  \label{Fv}
\end{equation}
where $\nu_{\rm R}$ is the $R$-band frequency, $E$ is given by Equation (\ref{Et}).

To constrain the parameters, we notice that (i) the R-band flux at $t = 19$ ks is
$F_{\nu_{\rm R}}\left(19 ~{\rm{ks}}\right)\simeq0.19$ mJy\footnote{This value has been corrected for Galactic and host galaxy extinction
with $E(B-V)=0.04$ mag \citep{Sch98} and $E(B-V)=0.02$ mag \citep{De16}, respectively.},
(ii) $\nu_{\rm c}$ should be well above the X-ray band at 2 days, i.e., $\nu_{\rm c}\left(2{~\rm{d}}\right)>10 ~{\rm{keV}}$,
(iii) $\nu_{\rm m}$ has crossed the $V$ band before 120 s, i.e., $\nu_{\rm m}\left(120~{\rm{s}}\right)<\nu_{\rm V}$.
Then we have
\begin{eqnarray}\label{con1}
E_{\rm f,52}^{1.52}\epsilon_{\rm e,-1}^{2.06}\epsilon_{\rm B,-2}^{1.02}n_{0}^{1/2} &\simeq& 4.1,\\ \label{con2}
E_{\rm f,52}^{-1/2}\epsilon_{\rm B,-2}^{-3/2}n_{0}^{-1}\left(1+Y\right)^{-2}  &>& 93.6, \\ \label{con3}
E_{\rm f,52}^{1/2}\epsilon_{\rm e,-1}^{2}\epsilon_{\rm B,-2}^{1/2} &<& 0.7,
\end{eqnarray}
where $Y$ is the Compton parameter denoting the energy ratio between the inverse Compton component and the synchrotron component. In the Thomson scattering regime, $Y=\left(\eta \epsilon_{\rm e}/\epsilon_{\rm B}\right)^{1/2}$ if $Y\gg1$, while $Y=\eta \epsilon_{\rm e}/\epsilon_{\rm B}$ if $Y\ll 1$, where
$\eta= {\rm min}\left\{1,\left(\nu_{\rm m}/\nu_{\rm c}\right)^{\left(p-2\right)/2}\right\}$ is the fraction of electron energy that was radiated away \citep{Sari01}.

From Equations (\ref{con1})$-$(\ref{con3}), one derives
\begin{eqnarray}
E_{\rm f,52}&>& 48\epsilon_{\rm e,-1}^{-1} \left(1+Y\right),\\
\epsilon_{\rm B,-2}&<&0.01 \epsilon_{\rm e,-1}^{-3} \left(1+Y\right)^{-1}.
\end{eqnarray}
We simply assume $E_{\rm i}= 2E_{\gamma}=2.1\times10^{52}$ erg. Since we have
$E_{\rm f}=E_{\rm i} \left(t_{\rm f}/t_{\rm i}\right)^{0.9}=33.8E_{\rm i}$, then we get
$E_{\rm f,52}=71$. For the other parameters, we take $\epsilon_{\rm e,-1}=1$ and
$\epsilon_{\rm B,-2}=5\times10^{-3}$, then $n_0\simeq2.0$ is obtained from Equation (\ref{con1}).
Using these parameters, we check that $Y<1$ is satisfied throughout the afterglow stage.
Our derived kinetic energy of $E_{\rm f}=7.1\times 10^{53}$ erg exceeds the usual
maximum rotational energy ($\sim 3\times 10^{52}$ erg) of a 1.4 $M_{\odot}$ magnetar. It leads to
the doubt that whether such a huge energy can be successfully supplied by the central engine.
A very massive magnetar may help to alleviate the difficulty, since the rotational energy could then
be up to $\sim 2 \times 10^{53}$ erg (Metzger et al. 2015). In the case of GRB 130831A, we argue
that the beaming effect may be a more realistic choice. When a beaming factor of
$f_{\rm b}>7.56\times10^{-3}$ \citep{De16} is considered, the intrinsic kinetic energy
will be reduced to $E_{\rm f,b}>5.4\times10^{51}$ erg, which is compatible with the usual magnetar energy limit.

Using the parameters given above, we can calculate the initial Lorentz factor by $\gamma_0\approx77.5 E_{\rm i,52}^{1/8} n_{0}^{-1/8} t_{2}^{-3/8} (1+z)^{3/8}$ \citep{Sari98}, and get $\gamma_0\approx95$. At the end of the energy injection, the Lorentz factor $\gamma=\gamma_0 \left(t_{\rm f}/t_{\rm i}\right)^{-0.26}\approx34$, then the ejecta distribution can be described by  $M\left(>\gamma\right)\propto \gamma^{-4.4}$, with $34\le\gamma\le95$.

\subsection{Afterglow of Internal Origin}\label{internal}
According to our fit, the X-ray afterglow LC shows an initial short plateau that lasts for $\sim$ 200 -- 300 s, followed by a steeper plateau between 0.3 and 98 ks, then it drops rapidly with a slope of $\sim$ 6. This steep internal plateau cannot be produced through the process of dipole spin-down of a magnetar.
For a rotating proto-neutron star, strong GW radiation could be produced through some nonaxisymmetric stellar perturbations, such as dynamical instabilities and
secular gravitational-wave driven instabilities (see \citet{Kok08} for a review). The latter are frame-dragging instabilities usually called Chandrasekhar-Friedman-Schutz instabilities \citep{Cha70,Fri78}, which is an efficient mechanism for the production of GWs and has a characteristic timescale compatible with the one of the GRB plateaus \citep{Corsi09}. As stated in Section \ref{intro}, when the neutron star undergoes GW-driven r-mode instability \citep{and98,Fri98}, GW losses can dominate the spin-down up to $\sim$ a few$\times10^6$ s \citep{Sa05}, so that the relatively steep and long internal X-ray plateau of GRB 130831A can be produced. In the following, we will evaluate the evolution of the magnetic dipole luminosity following the formalism of \citet{Yu10}.

According to  a phenomenological second-order model for the r-mode evolution \citep{Owen98,Sa04}, the spin of a magnetar evolves as \citep{Sa04,Yu09a}
\begin{equation}
\frac{dP}{dt}=\frac{4{\bar\alpha}^2}{15}\left(\delta+2\right)\frac{P}{\tau_{\rm g}}+\frac{P}{\tau_{\rm m}}, \label{P}
\end{equation}
where $P$ is the spin period of the magnetar, $\bar\alpha$ is the dimensionless amplitude of the r-modes, and $\delta$ is a free parameter describing the initial degree of the differential rotation of the star. The gravitational braking timescale can be written as $\tau_{\rm g}=37(P/P_{\rm K})^6$ s, where $P_{\rm K}$ is the Keplerian period at which the star starts shedding mass at the equator. The magnetic braking timescale is given by $\tau_{\rm m}=4\times10^5 I_{45} B_{14}^{-2} P_{-3}^{2}R_6^{-6}$ s, where $I_{45}$ is the moment of inertia in units of $10^{45}$ g cm$^2$ and $R_6$ is the radius of the magnetar in units of $10^6$ cm.
The evolution of the r-mode amplitude $\bar\alpha$ can be calculated from \citep{Sa04,Yu09a}
\begin{equation}
\frac{d\bar\alpha}{dt}=\left[1+\frac{2\bar{\alpha}^2}{15}\left(\delta+2\right)\right]\frac{\bar\alpha}{\tau_{\rm g}}+\frac{\bar\alpha}{2\tau_{\rm m}}. \label{alpha}
\end{equation}
In the case of $\tau_{\rm g,i}\ll\tau_{\rm m,i}$ (i.e., $B\ll B_{\rm c}=5\times10^{15} P_{{\rm i},-3}^{-2}$ G; the subscript ``i'' of a physical quantity means its initial value here and after), the spin-down would be first dominated by the gravitational wave radiation.  By ignoring the magnetic term, Equations (\ref{P}) and (\ref{alpha}) can be solved analytically, and $P(t)$ can be described by \citep{Sa05,Sa06}

\begin{equation}
P\left(t\right)\approx \left\{
 \begin{array}{ll} \label{Pt}
 P_{\rm i}\left[1-\frac{2}{15}\bar{\alpha}_{\rm i}^2 \left(\delta+2\right) {\rm exp}\left(2t/\tau_{\rm g,i}\right)\right]^{-1}, & {\rm for} ~ t<T_{\rm g},\\
 1.6P_{\rm i}\left(t/\tau_{\rm g,i}\right)^{1/5}, & {\rm for} ~ t>T_{\rm g},
 \end{array}
 \right.
 \end{equation}
where the break time $T_{\rm g}$ corresponds to the moment when the mode's amplitude changes from an exponential to a much slower power-law growth, and
can be solved from $d^2\bar \alpha/d t^2|_{t=T_{\rm g}}=0$ to be \citep{Yu09b}
\begin{eqnarray}
T_{\rm g}&=& -37\left[\ln \bar \alpha_{\rm i}+\frac{1}{2}\ln\left(\delta+2\right)+\frac{1}{2}\ln\left(\frac{6}{5}\right)\right]\left(\frac{P_{\rm i}}{P_{\rm K}}\right)^6 \nonumber\\
         &\equiv& T_{\rm K}\left(\bar\alpha_{\rm i},\delta\right)\left(\frac{P_{\rm i}}{P_{\rm K}}\right)^6.
\end{eqnarray}
Within a wide parameter region of $10^{-10}<\bar\alpha_{\rm i}<10^{-6}$ and $0<\delta<10^8$,  $T_{\rm K}$ varies from 170  to 840 s \citep{Sa05,Sa06,Yu09b}.
As the spin period increases, the magnetic braking effect would eventually exceed the gravitational braking effect and then $P$ evolves from $P\propto t^{1/5}$ to $P\propto t^{1/2}$. The transition happens at $T_{\rm c}$ given below.

We can calculate the magnetic dipole luminosity of a magnetar by \citep{Yu10}
 \begin{equation}
 L_{\rm md}\left(t\right)=10^{47} F\left(t\right) B_{14}^2 P_{{\rm i},-3}^{-4}R_6^6 ~~ {\rm erg ~s^{-1}},  \label{Lmd}
 \end{equation}
where for $B<B_{\rm c}$,
\begin{equation}\label{F}
F\left(t\right)\approx \left\{
\begin{array}{ll}
t^0, & t<T_g,\\
\left(\frac{t}{T_{\rm g}}\right)^{-q}, &  T_{\rm g}<t<T_{\rm c},\\
\left(\frac{T_{\rm c}}{T_{\rm g}}\right)^{-q}\left(\frac{t}{T_{\rm c}}\right)^{-2}, &  t>T_{\rm c}.
\end{array}
\right.
\end{equation}
Here $T_{\rm c}=\left(T_{\rm m}^2/T_{\rm g}^q\right)^{1/\left(2-q\right)}$ corresponds to the transition time when the magnetic braking effect exceeds the gravitational braking effect, and $T_{\rm m}=2\times10^5 I_{45}B_{14}^{-2} P_{\rm i,-3}^2 R_6^{-6}$ s is the initial spin-down timescale.
According to Equation (\ref{Pt}), $q$ can be taken as 0.8 approximately.

We note that the magnetic dipole emission does not convert completely  into X-rays.  By considering the converting effect and the correction for
the beaming of the magnetar wind, we can write the observed X-ray luminosity as
 \begin{equation}
 L_{\rm X}=\eta_{\rm X} L_{\rm md}/f_{\rm b}=10^{47} \eta_{\rm X,-1}f_{\rm b,-1}^{-1} L_{\rm md,47}~~ {\rm erg ~s^{-1}}, \label{Lx}
 \end{equation}
where $\eta_{\rm X}$  and $f_{\rm b}$ are the efficiency in converting the magnetic dipole emission into X-ray radiation and the beaming factor of the magnetar wind, respectively.

For GRB 130831A, however, we revise this scenario by requiring the magnetar collapses into a black hole well before $T_{\rm c}$ to explain the sharp drop of the X-ray LC at about $10^5$~s. To constrain the parameters, we require: (i) $\left(1+z\right)T_{\rm g}\simeq269$ s, (ii) the initial plateau luminosity $L_{X}\simeq2.6\times10^{47} {\rm erg~s^{-1}}$, (iii) $\left(1+z\right)T_{\rm c}>9.8\times10^4$ s.
We assume $T_{\rm K}=170$ s and $\eta_{X}/f_{\rm b}=1$, then following the expressions of $T_{\rm g}$, $T_{\rm c}$ and Equations (\ref{Lmd})$-$(\ref{Lx}), we obtain the initial spin period $P_{\rm i}\simeq1.01P_{\rm K}$ and the  magnetic field strength $B_{14}\simeq1.0$.
We note that our obtained spin period is very close to the Keplerian period (i.e., the minimum spin period allowed before breakup), which can be taken as 0.8 ms for a 1.4 $M_{\odot}$ neutron star \citep{Lat04,Hae09}.

\section{COMPARISON WITH OBSERVED LIGHT CURVES}\label{fitting}
We have calculated the theoretical multi-band afterglow LCs based on
Equations (\ref{Et}), (\ref{Fv}), (\ref{Lmd})$-$(\ref{Lx}) and on our derived parameters
in Section \ref{model}. For the decay slope after the magnetar collapses into
a black hole, we artificially set it as 6.8 according to the fitting results of \citet{De16}.

\begin{figure}
\includegraphics[scale=.50]{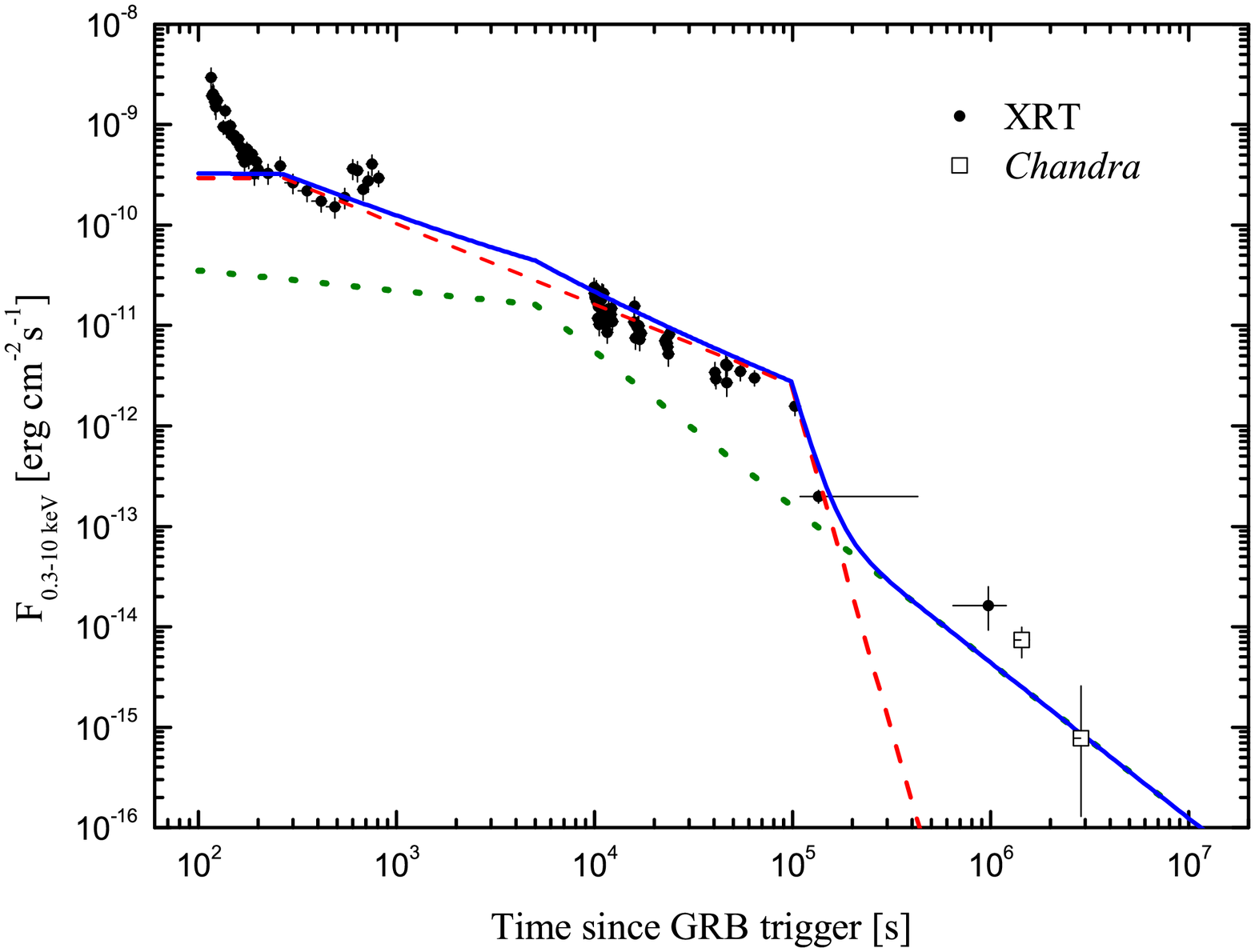}
\caption{Our theoretical X-ray afterglow light curves as compared with
the observed X-ray flux of GRB 130831A. The parameter values of $\epsilon_{\rm e,-1}=1$, $\epsilon_{\rm B,-2}=5\times10^{-3}$,
$E_{\rm f,52}=71$, $n_0=2.0$, $P_{\rm i,-3}=0.8$ and $B_{14}=1.0$ are used. The XRT data (dots) are taken from
http://www.swift.ac.uk/burst\_analyser/00568849/ \citep{Evans09}. The {\it Chandra} data (squares) are taken from
\citet{De16}. The dotted and dashed lines correspond to calculated external and internal afterglow components, respectively.
The solid line is the superposition of both components. The initial steep decay and X-ray
flare are not considered. \label{Xfit2}}
\end{figure}

\begin{figure}
\includegraphics[scale=.50]{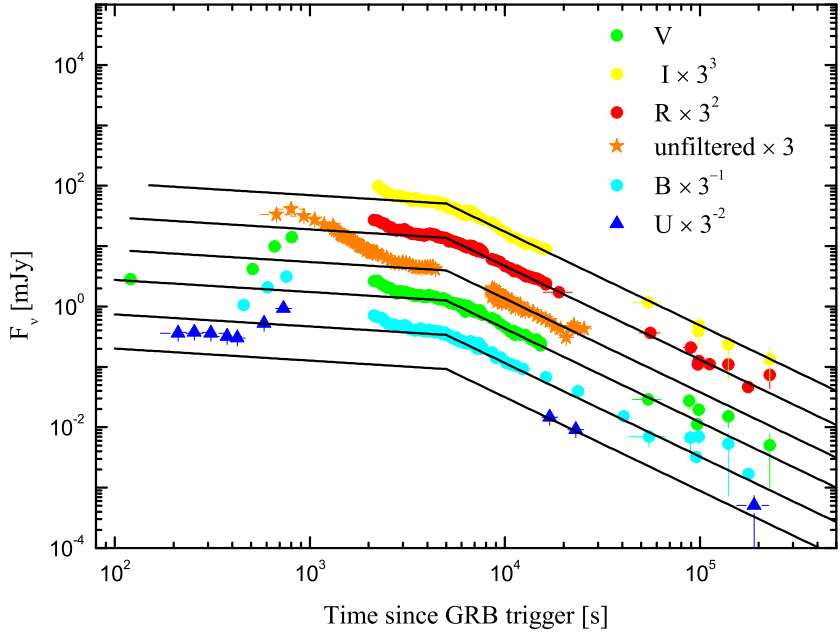}
\caption{Our theoretical multi-band optical afterglow light curves as compared with
the observations of GRB 130831A. The parameter values of $\epsilon_{\rm e,-1}=1$, $\epsilon_{\rm B,-2}=5\times10^{-3}$,
$E_{\rm f,52}=71$, $n_0=2.0$ are used. The predicted emissions are from the FS (solid lines). The
data are taken from \citet{De16} and have been corrected for Galactic and host galaxy extinction.
Adjacent light curves have been offset by a factor of three for clarity, with $V$ band unscaled.
The optical flares are not considered.\label{optfit2}}
\end{figure}

Figure~\ref{Xfit2} compares our theoretical 0.3 -- 10 keV LC with the X-ray flux observed by XRT and {\it Chandra}.
It is shown that our model can basically describe the observed flux evolution. We note, however,
our model slightly overestimates the internal plateau. This is because we use the plateau decay
index of 0.8 following Equations (\ref{Pt}) and (\ref{F}) which are solved  analytically, a
slightly larger value ($\sim$ 1) can be obtained from a more accurate numerical calculation \citep{Yu10},
which would improve the modeling. Moreover, the external afterglow has a small contribution
to the internal X-ray plateau. In addition, the theoretical X-ray flux underestimates the
late XRT and {\it Chandra} data points except for the last one which has very large errors.
This is to be expected, since the theoretically predicted decay index of the FS component
is 1.55, which is slightly steeper than the observed value ($\sim1.1$). However, when the uncertainties are considered,
they are consistent at 2.1$\sigma$ CL \citep{De16}.
For the multi-band optical afterglow plot in Figure \ref{optfit2}, our model also gives a
good explanation except for the $U$ band data, which show a slight excess between 200 and 500 s.
The reason may be that the energy injection model adopted here is still too simplified. More realistic
and complicated external shock processes might be involved and need further investigations.

\begin{figure}
\includegraphics[scale=.50]{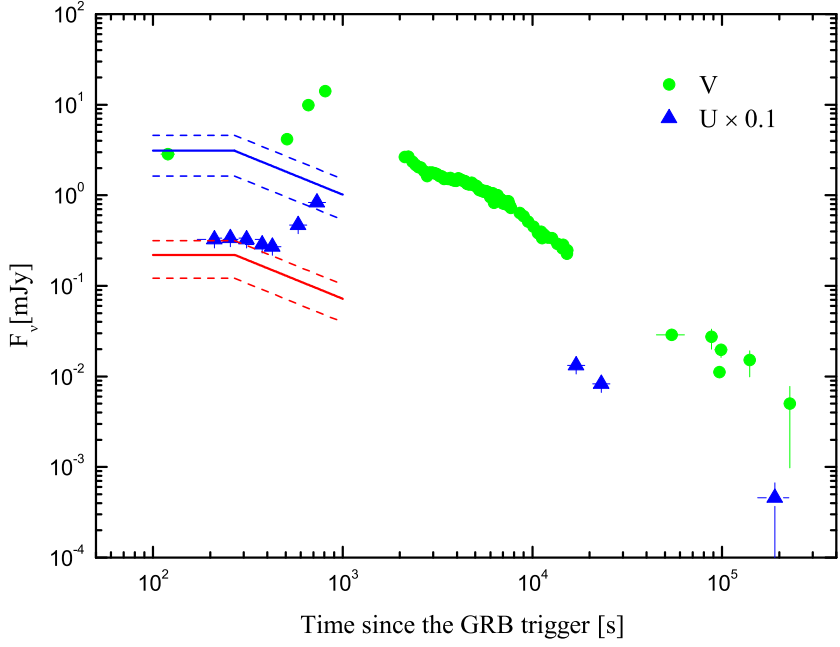}
\caption{Predicted $U$ and $V$ band afterglow (solid lines) from the  magnetar wind internal
emission as compared with the observed early afterglow. The dashed lines correspond to
the 1$\sigma$ CL. The data are taken from \citet{De16} and have been corrected for Galactic
and host galaxy extinction. The $U$ band data have been rescaled by a factor of 0.1 for clarity.
\label{Ufit}}
\end{figure}

Another possibility is that the magnetar wind emission may have a significant contribution to the initial optical plateau.
According to our magnetar model and our fitting results of the X-ray afterglow, the wind emission could produce a short X-ray plateau lasting for $\sim$ 300 s, which is compatible with the timescale of the early optical plateau. To check the flux level, we assume that the magnetar wind emission can extend to the optical band with the  spectral index $\beta_{\rm X}=0.77\pm0.07$, then extrapolate the flux density of the initial X-ray  plateau to the optical band using the fitted flux parameters (see Table \ref{table1}), and compare with the early data in the $U$ and $V$ bands. As shown in Figure \ref{Ufit}, the predicted initial optical plateau is now in accord with the $V$ band data rather well, and it is also consistent with the $U$ band data at 1$\sigma$ CL. In this case, even the energy injection from the refreshed shocks is no longer needed for this short period. The optical afterglow after $\sim$ 3 ks can be interpreted in the context of the standard afterglow model, and the late plateau (between $\sim$ 3 and 5 ks) is produced by the superposition of a decaying optical flare and a rising FS peak \citep{De16}. However, to avoid exceeding the external component at late times ($\sim$ 100 ks), we have to assume that the internal optical emission fade rapidly at some time. Moreover, according to this explanation,
the flux before and after the flare have roughly the same normalization but with completely different origins, which makes this scenario somewhat contrived.

\section{CONCLUSIONS AND DISCUSSIONS} \label{conclusion}
The X-ray afterglow of GRB 130831A shows a shallow decay  followed by a steep drop at about $10^5$ s, which is the signature of an ``internal plateau''.
After the drop, the X-ray afterglow continues with a much shallower decay, a similar decay behavior appears in the optical afterglow after $\sim$ 5 ks. Before $\sim$ 5 ks, the optical afterglow shows two segments of plateaus separated by a pronounced optical flare peaking at $\sim$ 730 s.
The decay slope of the internal X-ray plateau ($\sim$ 0.8) is so steep that it cannot be explained by the simplest magnetar spin-down model. We interpret this special internal plateau as the magnetar wind  emission during the spin-down process dominated by GW radiation losses in the context of r-mode instability, and explain the steep drop by assuming that the magnetar collapses into a black hole. This scenario also predicts an initial X-ray plateau lasting for hundreds  of seconds which is compatible with observations.
By assuming that the two segments of optical plateaus have the same external origin and the magnetar wind emission has a negligible contribution in the optical band, we interpret the optical and late X-ray afterglow as FS emission by invoking the energy injection from a continuously refreshed shock following the prompt emission phase. It is shown that the U-band afterglow has a slight excess before $\sim$ 500s. One possibility is that our analytical treatment of the energy injection model is still too simplified and more complicated external shock processes might be called for. Another solution is that the magnetar wind emission
might have a significant contribution to the early optical afterglow.

The observed X-ray internal plateaus are typically flat and can be produced through the spin-down process caused by dipolar radiation before the magnetar collapses into a black hole. About 29$-$56\% short GRBs exhibit an internal plateau \citep{Row13,Lu15,Gao16}, whereas this is much lower for long GRBs \citep[10 candidates identified in][]{Lyons10}. The magnetic field strength and spin period required to reproduce the observed internal plateaus in both long and short GRBs are typically $P_{\rm i}\sim 1-10$ ms and $B\sim10^{15}-10^{16}$~G \citep{Lyons10,Row13,Lu15}.
For GRB 130831A, our derived parameters of $P_{\rm i}\simeq0.8$ ms and $B\simeq10^{14}$ G are at the lower end of these distributions. This is not surprising, however, since the internal plateau of this burst is quite different from those usually studied. In our magnetar model, the extreme spin period is solely determined by the break time $T_{\rm g}\sim270$ s, while the weak magnetic field strength is determined by the luminosity of the X-ray initial plateau. Besides, such a weak magnetic field strength is also required by the long duration of the internal plateau of GRB 130831A. To produce this plateau, the GW losses would dominate the spin-down until $\sim10^5$ s (i.e., $(1+z)T_{\rm c}>10^5$ s), which gives $B_{14}<4$. Therefore, it is the unique feature of the internal plateau of GRB 130831A that determines the extreme spin period and magnetic field strength.

In our magnetar scenario, the magnetic dipole radiation is strong and significant up to $\sim$ 100 ks,
and it can successfully power the internal X-ray plateau. It is assumed that this dipole radiation component
mainly contributes to the emission in X-ray band, but contribute negligibly at optical wavelengths. To
interpret the optical plateau that lasts for $\sim$ 5 ks in the case of GRB 130831A, we invoke the energy
injection from a continuously refreshed shock following the prompt emission phase. We would like to stress
that the hypothesis that the magnetic dipole radiation mainly contributes to X-ray emission but not optical
emission is still an assumption. It needs to be further checked by future observations.
Interestingly, it seems that the assumption already received some support from observations.
Previous studies have indicated that the magnetar wind emission tends to contribute to the internal plateau
mostly in X-rays, but it is usually not detected at optical bands \citep{Troja07,Lyons10,Row13}.
\citet{Row13} found that several short GRBs with internal plateaus have optical afterglows which are
consistent with their X-ray afterglows during the plateau phase, but these cases would require some
extreme parameters. Therefore, whether the magnetar wind emission produces the internal plateau in the
optical band is not conclusive in their analysis.

To produce the multi-band afterglow of GRB 130831A, our model invokes two different outflow components.
One is an ``active'' outflow, which internally produces X-ray emission for $\sim$ 100 ks. The other outflow, instead,
does not produce internal emission, but still increases the energy of the leading ejecta for $\sim$ 5 ks, leading to
the observed optical plateau. This scenario is somewhat more complicated as compared with many previous GRB scenarios
in which usually only one component is involved. Since the multi-band afterglow behavior of GRB 130831A itself is
very complicated, we believe that such a choice is still reasonable.
We argue that in our scheme, the external and internal plateaus may coexist in the multi-band afterglow.
It can be tested by more similar observations in the future.

\acknowledgments

We acknowledge the anonymous referee for helpful comments and suggestions. This work made use of data supplied by the UK Swift Science Data Centre at the University of Leicester. Our study was supported by the National Basic Research Program of China with Grant No. 2014CB845800  and by the National Natural Science Foundation of China with Grants No. 11473012, No. 11475085, and No. 11275097.

\clearpage

\begin{deluxetable}{cccccccccc}
\tabletypesize{\scriptsize}
\rotate
\tablecaption{Fitting parameters of the X-ray afterglow before $10^5$ s.  \label{table1}}

\tablewidth{0pt}
\tablehead{
\colhead{Model} & \colhead{$\alpha_{\rm PL}$} & \colhead{$f_0$}  & \colhead{$\alpha_1$} & \colhead{$t_{\rm b}$} &
\colhead{$\alpha_2$} & \colhead{$f_1$} & \colhead{$t_{\rm c}$} & \colhead{$w$} & \colhead{$\chi^2/dof$} \\
\colhead{} & \colhead{} & \colhead{($\times10^{-10}$ erg cm$^{-2}$ s$^{-1}$)} & \colhead{} & \colhead{(s)}&
\colhead{} & \colhead{($\times10^{-10}$ erg cm$^{-2}$ s$^{-1}$)} & \colhead{(s)} & \colhead{(s)} & \colhead{}
}

\startdata
PL+BPL+Gaussian & 5.52$\pm$ 0.92 & 2.93$\pm$0.62 & 0 (fixed) & 269$\pm$75 & 0.85$\pm$0.02  & 1.94$\pm$0.37 & 760$\pm$47  & 165$\pm$66 & 76/80 \\
\enddata
\tablecomments{The fitting functions are: (a) a PL with the decay index $\alpha_{\rm PL}$;
(b) a BPL with the decay indices $\alpha_1$, $\alpha_2$,  break time $t_{\rm b}$ and flux normalisation $f_0$, we fix $\alpha_1=0$;
(c) a Gaussian function presented by $f=f_1 {\rm exp} \left[-(t-t_{\rm c})^2/w^2\right]$.}
\end{deluxetable}

\begin{deluxetable}{llllllll}
\tabletypesize{\scriptsize}
\tablecaption{Fitting parameters of the  optical afterglow before 16 ks.  \label{table2}}
\tablewidth{0pt}
\tablehead{
\colhead{Filter} & \colhead{$\alpha_1$} & \colhead{$t_{\rm b}$}  & \colhead{$\alpha_2$} &
\colhead{$\alpha_{1}^{\rm F}$} & \colhead{$t_{\rm b}^{\rm F}$} & \colhead{$\alpha_{2}^{\rm F}$} & \colhead{$\chi^2/dof$} \\
\colhead{} & \colhead{} & \colhead{(ks)} & \colhead{} &
\colhead{} & \colhead{(ks)} & \colhead{} &\colhead {}
}
\startdata
$V$ & 0.22$\pm$0.06 & 5.51$\pm$0.14 & 1.55$\pm$0.04 & -4.9$\pm$1.2 & 0.75$\pm$0.03 & 2.5$\pm$0.1 & 254/74 \\
$B$ & -1.79$\pm$0.62 & 4.66$\pm$0.11 & 1.32$\pm$0.06 & -2.4$\pm$0.2 & 0.75 (fixed) & 1.6$\pm$0.1 & 105/62 \\
unfiltered & -0.97$\pm$0.37 & 4.77$\pm$0.17 & 1.51$\pm$0.08 & -3 (fixed) & 0.81$\pm$0.02 & 2.2$\pm$0.1 & 197/99 \\
$R$ & -0.47$\pm$0.17 & 4.96$\pm$0.10 & 1.40$\pm$0.03 &   &   & 2.3 (fixed) & 284/74 \\
$I$ & -0.79$\pm$0.27 & 4.75$\pm$0.11 & 1.35$\pm$0.05 &   &   &  2.3 (fixed) & 208/65 \\
$U$ & 0.29$\pm$0.10 &                &              &    &   &             & 8/3 \\

\enddata
\tablecomments{The $V, B$ and unfiltered bands are fitted with two BPLs, while the $R$ and $I$ bands are fitted with a PL+BPL. For the $U$ band, we fit only the initial plateau with a PL. The temporal parameters of the plateau components are $\alpha_1$, $\alpha_2$ and $t_{\rm b}$, while those with the superscript ``F'' correspond to the optical flares.}
\end{deluxetable}

\end{document}